\documentclass[a4paper,fleqn,usenatbib]{mnras}

\usepackage{txfonts}

\usepackage[T1]{fontenc}
\usepackage{ae,aecompl}

\usepackage{verbatim}
\usepackage{gensymb}
\usepackage{graphicx}
\usepackage{array}
\usepackage{booktabs}
\usepackage{threeparttable}
\usepackage{natbib}

\newcommand{\otoprule}{\midrule[\heavyrulewidth]}
\newcolumntype{+}{>{\global\let\currentrowstyle\relax}}
\newcolumntype{^}{>{\currentrowstyle}}

\makeatletter
\newlength{\abovecaptionskip}%
\makeatother

\title[Short title, max. 45 characters]{Non-Detection of HC$_{11}$N toward TMC-1: Constraining the Chemistry of Large Carbon-Chain Molecules}

\author[R. A. Loomis et al.]{
Ryan A. Loomis,$^{1}$\thanks{E-mail: rloomis@cfa.harvard.edu}
Christopher N. Shingledecker,$^{2}$
Glen Langston,$^{3}$
\newauthor
Brett A. McGuire,$^{4,5,8}$
Niklaus M. Dollhopf,$^{6,9}$
Andrew M. Burkhardt,$^{6}$
Joanna Corby,$^{6}$
\newauthor
Shawn T. Booth,$^{6,10}$
P. Brandon Carroll,$^{7}$
Barry Turner,$^{4,\dagger}$
and
Anthony J. Remijan$^{4}$
\\
$^{1}$Department of Astronomy, Harvard University, Cambridge, MA 02138\\
$^{2}$Department of Chemistry, University of Virginia, Charlottesville, VA 22904\\
$^{3}$National Science Foundation, Division of Astronomical Sciences, Arlington, VA 22230\\
$^{4}$National Radio Astronomy Observatory, Charlottesville, VA 22904\\
$^{5}$Harvard-Smithsonian Center for Astrophysics, Cambridge, MA 02138\\
$^{6}$Department of Astronomy, University of Viriginia, Charlottesville, VA 22904\\
$^{7}$Division of Chemistry and Chemical Engineering, California Institute of Technology, Pasadena, CA 91125\\
$^{8}$B.A. McGuire is a Jansky Fellow of the National Radio Astronomy Observatory\\
$^{9}$N. Dollhopf was a summer student at the National Radio Astronomy Observatory\\
$^{10}$S.T. Booth was a summer student at the National Radio Astronomy Observatory
}

\date{Accepted 8 September 2016}

\pubyear{2016}

\begin{document}
\label{firstpage}
\pagerange{\pageref{firstpage}--\pageref{lastpage}}
\maketitle

\begin{abstract}
\cite{Bell_1997} reported the first detection of the cyanopolyyne HC$_{11}$N toward the cold dark cloud TMC-1; no subsequent detections have been reported toward any source.  Additional observations of cyanopolyynes and other carbon-chain molecules toward TMC-1 have shown a log-linear trend between molecule size and column density, and in an effort to further explore the underlying chemical processes driving this trend, we have analyzed GBT observations of HC$_9$N and HC$_{11}$N toward TMC-1.  Although we find an HC$_9$N column density consistent with previous values, HC$_{11}$N is not detected and we derive an upper limit column density significantly below that reported in \cite{Bell_1997}.  Using a state-of-the-art chemical model, we have investigated possible explanations of non-linearity in the column density trend.  Despite updating the chemical model to better account for ion-dipole interactions, we are not able to explain the non-detection of HC$_{11}$N, and we interpret this as evidence of previously unknown carbon-chain chemistry. We propose that cyclization reactions may be responsible for the depleted HC$_{11}$N abundance, and that products of these cyclization reactions should be investigated as candidate interstellar molecules.
\end{abstract}




\section{Introduction}

{\let\thefootnote\relax\footnote{$\dagger$ Deceased}}Carbon chains are the starting point for a significant fraction of the known chemical complexity within the interstellar medium \citep{Thaddeus_2001}. In addition to forming the backbone of the polyyne families (cyanopolyynes, methylpolyynes, and methylcyanopolyynes) \citep[e.g.][]{Irvine_1981, Bell_1997, Snyder_2006, Remijan_2006}, carbenes such as H$_2$C$_5$ and H$_2$C$_6$ \citep{McCarthy_1997_Sci, McCarthy_1997_8mol}, and unsaturated hydrocarbons such as HC$_4$H and HC$_6$H \citep{Cernicharo_2001}, they are also the precursors of a number of interstellar anions \citep{McCarthy_2006, Brunken_2007, Cernicharo_2007, Thaddeus_2008}. Additionally, carbon-chain species may play an important role in the formation of polycyclic aromatic hydrocarbons (PAHs) \citep{Tielens_2008, Duley_2009}, and are promising candidates as carriers of the diffuse interstellar bands \citep{Thaddeus_1995, Tulej_1998, Motylewski_2000, Thaddeus_2001, Maier_2004, Jochnowitz_2008, Zack_2014}.  Thus, studying the formation and destruction chemistry of carbon-chain molecules promises to provide insight into a large subset of interstellar chemistry.

A particularly well-studied family of carbon-chains are the cyanopolyynes: linear molecules of the form $\mathrm{HC_\textit{n}N}$, where $n=3,5,7,9, etc.$ As with the other carbon-chains, cyanopolyynes are observed in high abundance toward asymptotic giant branch (AGB) stars \citep[e.g.][]{Truong-Bach_1993, Agundez_2008} and cold dark clouds such as the Taurus molecular cloud (TMC-1) \citep[e.g.][]{Bujarrabal_1981, Hirahara_1992, Ohishi_1998, Kaifu_2004, Gratier_2016}. The polyyne families of molecules in TMC-1 share a surprisingly linear correlation between their log abundances and size \citep{Bujarrabal_1981, Bell_1997, Ohishi_1998, Remijan_2006}, which can be explained by a formation chemistry governed by a small set of gas-phase reactions to add carbons to the chain \citep{Winnewisser_1979, Bujarrabal_1981, Fukuzawa_1998}.  Observations of AGB stars such as IRC+10216 and CRL2688 also show abundances of cyanopolyynes that follow the same downward trend \citep[e.g.][]{Truong-Bach_1993}, suggesting that the underlying gas-phase formation mechanisms are similar for both AGB stars and cold cores.

Given this linear trend and assumption of relatively simple chemistry, both laboratory and observational studies have attempted to find larger cyanopolyynes. In the laboratory, cyanopolyyne chains up to HC$_{17}$N have been detected \citep{McCarthy_1998_HC17N}, while HC$_{11}$N is the largest detected interstellar cyanopolyyne. Originally thought to be detected in the AGB star IRC+10216 \citep{Bell_1982}, refinements to transition rest frequencies of HC$_{11}$N brought the detection into dispute \citep{Travers_1996}, and a further search of IRC+10216 yielded a non-detection (M. Bell, unpublished).  Utilizing new laboratory data, a deep search of TMC-1 was undertaken by \cite{Bell_1997} using the NRAO 140 foot telescope at Green Bank, resulting in a reported detection of two HC$_{11}$N transitions roughly in agreement with predictions of the linear abundance trend. No subsequent detections of HC$_{11}$N have been reported toward any astronomical source.

In this paper, we present an analysis of observations originally taken with the intent to confirm the HC$_{11}$N detection, and to identify HC$_{13}$N toward TMC-1. We report a non-detection of HC$_{11}$N in conflict with the column density reported by \cite{Bell_1997}, and attempt to resolve these discrepant observations.  Utilizing a state-of-the-art chemical modeling code, we calculate cyanopolyyne abundances in TMC-1 and comment on possible reasons for a depleted abundance of HC$_{11}$N.

In Section 2 we describe the observations, in Section 3 the observational results are presented, and in Section 4 the observational method is reviewed for possible explanations of the HC$_{11}$N non-detection. In Section 5 we present our chemical modeling results and compare to the observations, and in Section 6 we summarize our results.

\section{Observations}
We utilize observations on the Robert C. Byrd Green Bank Telescope (GBT) from project AGBT06A\_046 (PI: G. Langston). Extensive details of the observations and calibration are presented in \cite{Langston_2007}, and are briefly summarized here. Observations of the TMC-1 cyanopolyyne peak (4$^h$41$^m$42$^s$, 25$\degr$ 41$\arcmin$ 27$\arcsec$ J2000) were taken in Ku-band (11.7 GHz to 15.6 GHz) over 28 epochs in 2006, for a total of $\sim$180 hrs integration time. For the first 12 epochs, the GBT spectrometer was configured for dual beam, dual polarization observations with four 12.5 MHz spectral windows. Each 12.5 MHz spectrum consisted of 4096 channels (3052 Hz resolution, 0.070 km s$^{-1}$).  In the remaining 16 epochs, an identical setup was used, but with a larger bandwidth of 50 MHz.  Each 50 MHz spectrum consisted of 16384 channels, providing an identical spectral resolution to the previous epochs.

Observations were taken using the ``nodding" mode, alternating between the two Ku-band receiver feeds, separated by an angular offset of 330\arcsec~on the sky. This angular separation is sufficiently large such that there is no significant cyanopolyyne emission from TMC-1 in the ``off" position feed. To confirm correct pointing, the HC$_{11}$N observations were preceded most days by short test observations of the HC$_5$N J=5--4, HC$_7$N J=11--10, HC$_7$N J=12--11, and HC$_9$N J=23--22 transitions. As reported by \cite{Langston_2007}, the observed HC$_7$N J=11--10 intensity was a factor of 2.05$\pm$.05 stronger than the value reported by \cite{Bell_1997}, suggesting both an accurate pointing as well as a beam filling factor of close to unity for the GBT observations (see Section 4).

Peak and focus observations were performed on the quasar 3C123. The data were reduced using the {\it rt-idl} and {\it GBTIDL} packages. Antenna temperatures were recorded on the T$_A$ scale with conservatively-estimated 20\% uncertainties, and were then converted to the T$_b$ scale by correcting for the main beam and aperture efficiencies (for a detailed discussion of the antenna temperature calibration, see \cite{Langston_2007}). A systemic velocity of v$_{LSR}$ = +5.8 km s$^{-1}$ was assumed \citep{Kaifu_2004}.

\section{Observational Results}
Six consecutive transitions of HC$_9$N and six consecutive transitions of HC$_{11}$N were covered by the observations.  Rest frequencies for all transitions were obtained from the Splatalogue database\footnote{Available at www.splatalogue.net \citep{Remijan_2007}.}\footnote{Frequencies are catalogued through CDMS \citep{CDMS_2001, CDMS_2005}.}. We are highly confident in these rest frequencies, as the original HC$_{11}$N laboratory data span the range of the astronomical observations (6--15 GHz), and the rest frequencies were measured to very high spectral resolution, i.e. \textless 0.1 km s$^{-1}$ \citep{Travers_1996, McCarthy_1997_8mol}. Additionally, $^{13}$C and $^{15}$N isotopic spectroscopy confirmed the elemental composition and geometry of the molecule \citep{McCarthy_2000}.

All six HC$_9$N transitions were detected at the systemic velocity of +5.8 km s$^{-1}$ (Figure 1), and their intensities and line-widths are reported in Table 1. A column density of 2.3$\pm$0.2 $\times$ 10$^{12}$ cm$^{-2}$ and rotational temperature of 10$\pm$2 K were calculated using the formalism described in \cite{Hollis_2004}, assuming optically thin emission that fills the telescope beam. These values are consistent with previous observations of HC$_9$N in TMC-1, as are the measured line-widths \citep{Bell_1997, Kaifu_2004, Kalenskii_2004, Remijan_2006}. We take this as evidence that there are no large systematic errors with the observations, and that our assumptions about both the beam filling factor and the ``off" nodding positions are reasonable (see Section 4).

\begin{figure}
\centering
\includegraphics[width=\columnwidth]{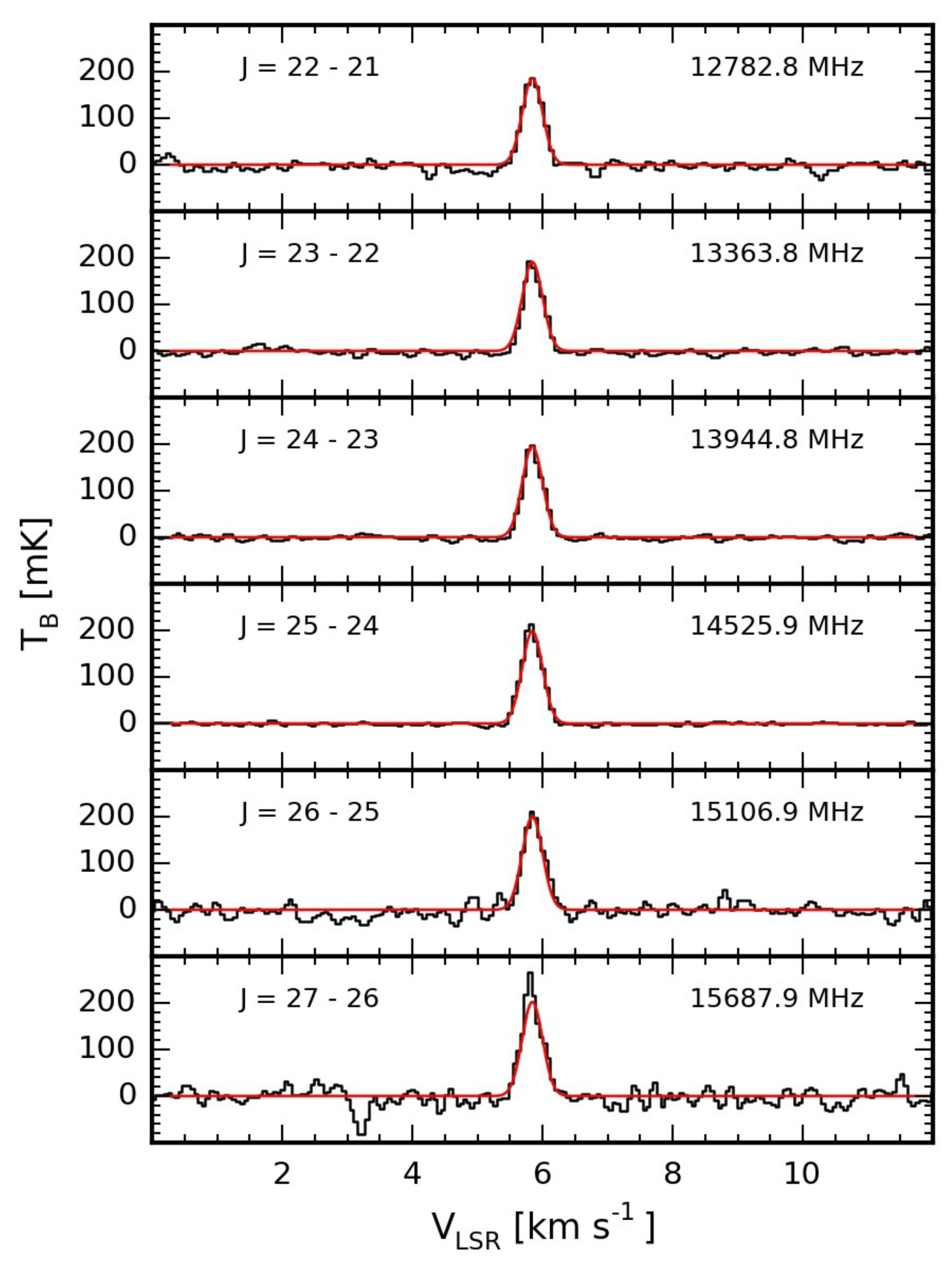}
\caption{{\small Observations of HC$_{9}$N transitions with predictions using the best fit column density and rotational temperature overplot in red.}}
\end{figure}

\begin{table*}
\begin{center}
\small
\begin{threeparttable}[b]
\caption{Observed transitions of HC$_{9}$N and HC$_{11}$N upper limits}
\begin{tabular}{+c^c^c^c^c^c}

    \multicolumn{6}{c}{\textbf{HC$_{9}$N}}\\
\toprule
	Transition & Frequency$^{ab}$ & E$_{u}$ & S$_{ij}\mu^{2}$ &  $\Delta$T$_{b}$  & $\Delta$V \\
	J' -- J"       & (MHz)         & (K)       & (D$^{2}$) & (mK) & (km s$^{-1}$) \\ 
	\otoprule
	22 -- 21     &	12782.8     &	7.05	&	595 	&	189(6)    &   0.37(2)   \\
	23 -- 22     &	13363.8     &	7.70	&	622 	&	188(4)    &   0.35(1)   \\
	24 -- 23     &	13944.8     &	8.37	&	649 	&	191(2)    &   0.36(1)   \\
	25 -- 24     &	14525.9     &	9.06	&	676     &	201(2)    &   0.39(1)   \\
    26 -- 25     &	15106.9     &	9.79	&	703 	&	201(9)    &   0.38(2)   \\
	27 -- 26     &	15687.9     &	10.54	&	730     &	226(11)   &   0.39(2)   \\
	\bottomrule
\\
\\
    \multicolumn{6}{c}{\textbf{HC$_{11}$N}}\\
\toprule
	Transition & Frequency$^{ab}$ & E$_{u}$ & S$_{ij}\mu^{2}$ & $\Delta$T$_{b}^{c}$  & Column Density\\
	J' -- J"       & (MHz)         & (K)       & (D$^{2}$) & (mK)  & (10$^{11}$ cm$^{-2}$) \\ 
	\otoprule
	38 -- 37     &	12848.7     &	12.02	&	1137	&	$<$ $\space$18.3     &	$<$ $\space$4.1     \\
	39 -- 38     &	13186.9     &	12.66	&	1167	&	$<$ $\space$6.8     &   $<$ $\space$1.5     \\
	40 -- 39     &	13525.0     &	13.31	&	1197	&	$<$ $\space$6.8	    &   $<$ $\space$1.5     \\
	41 -- 40     &	13863.1     &	13.97	&	1227	&	$<$ $\space$8.6	    &   $<$ $\space$2.0     \\
	42 -- 41     &	14201.2     &	14.65	&	1256	&	$<$ $\space$4.9     &   $<$ $\space$1.1     \\
	43 -- 42     &	14539.3     &	15.35	&	1287	&	$<$ $\space$5.2     &   $<$ $\space$1.2     \\
	\bottomrule
\end{tabular}
\begin{tablenotes}
\item[a] Beam sizes and efficiencies for each respective frequency can be found in \cite{Hollis_2007}.
\item[b] Rest frequencies were obtained from the Splatalogue database, see Sect. 3 for complete references.
\item[c] Brightness temperature upper limits are the 95\% confidence level values derived for each respective spectrum.
\end{tablenotes}
\end{threeparttable}
\end{center}
\end{table*}

Our observations of the six HC$_{11}$N transitions are shown in Figure 2; none were identified above a 2$\sigma$ significance at +5.8 km s$^{-1}$.  All of these transitions were high J (\textgreater \space 30) and $a$-type ($\mu_a$ = 5.47(5)~D) (P. Botschwina, private communication (1997) in \cite{Bell_1997}), similar to and including both transitions from \cite{Bell_1997}. The rms noise of the spectra ranges from 2--3 mK for four of the transitions, and 6--7 mK for the remaining two transitions, which had less integration time. The 3$\sigma$ noise levels for each respective spectrum are shown as dashed blue lines. Based on the column density reported in the \cite{Bell_1997} detection, and the characteristic cyanopolyyne properties of TMC-1 \citep[e.g.][]{Kaifu_2004}, we over-plot in red a predicted spectrum for each transition, using a line width of 0.4 km s$^{-1}$ and excitation temperature of 10 K \citep{Bell_1997}. These predicted spectra take into account the different beam filling factor of the cyanopolyyne emission region for the GBT and the NRAO 140ft telescope (see Section 5.1). The predicted intensities range from 11--13 mK, equivalent to a 4--5$\sigma$ signal in the higher SNR spectra, but we do not observe any signals of this strength.

To further constrain the abundance of HC$_{11}$N, we weighted the spectra by their variances and stacked them, shown in Figure 3. Spectral stacking is particularly well suited for the linear cyanopolyynes \citep[e.g.][]{Langston_2007}, which have regular and well characterized transition rest frequencies and smoothly varying intensities (e.g. Figure 1 and Table 1 for HC$_9$N). At an excitation temperature of 10 K, our predicted intensities only change by $\sim$15$\%$ from the least to most energetic observed transition. As noted by \cite{Bell_1998} and \cite{Ohishi_1998}, however, radiative cooling likely plays an important role in determining the cyanopolyyne excitation temperatures in TMC-1, with the smaller cyanopolyynes being more affected. Due to its size, HC$_{11}$N should be less affected, and \cite{Bell_1998} predicts an excitation temperature of 8--10 K. Even at a pessimistic excitation temperature of 5 K, however, our predicted intensities would only change by $\sim$33$\%$ from the least to most energetic observed transition.  Thus we believe that spectral stacking is a valid method for placing limits on HC$_{11}$N emission. The stacked HC$_{11}$N spectrum has an rms of 1.4 mK, and we observe no signal in excess of 2$\sigma$ at +5.8 km s$^{-1}$.

\begin{figure}
\centering
\includegraphics[width=\columnwidth]{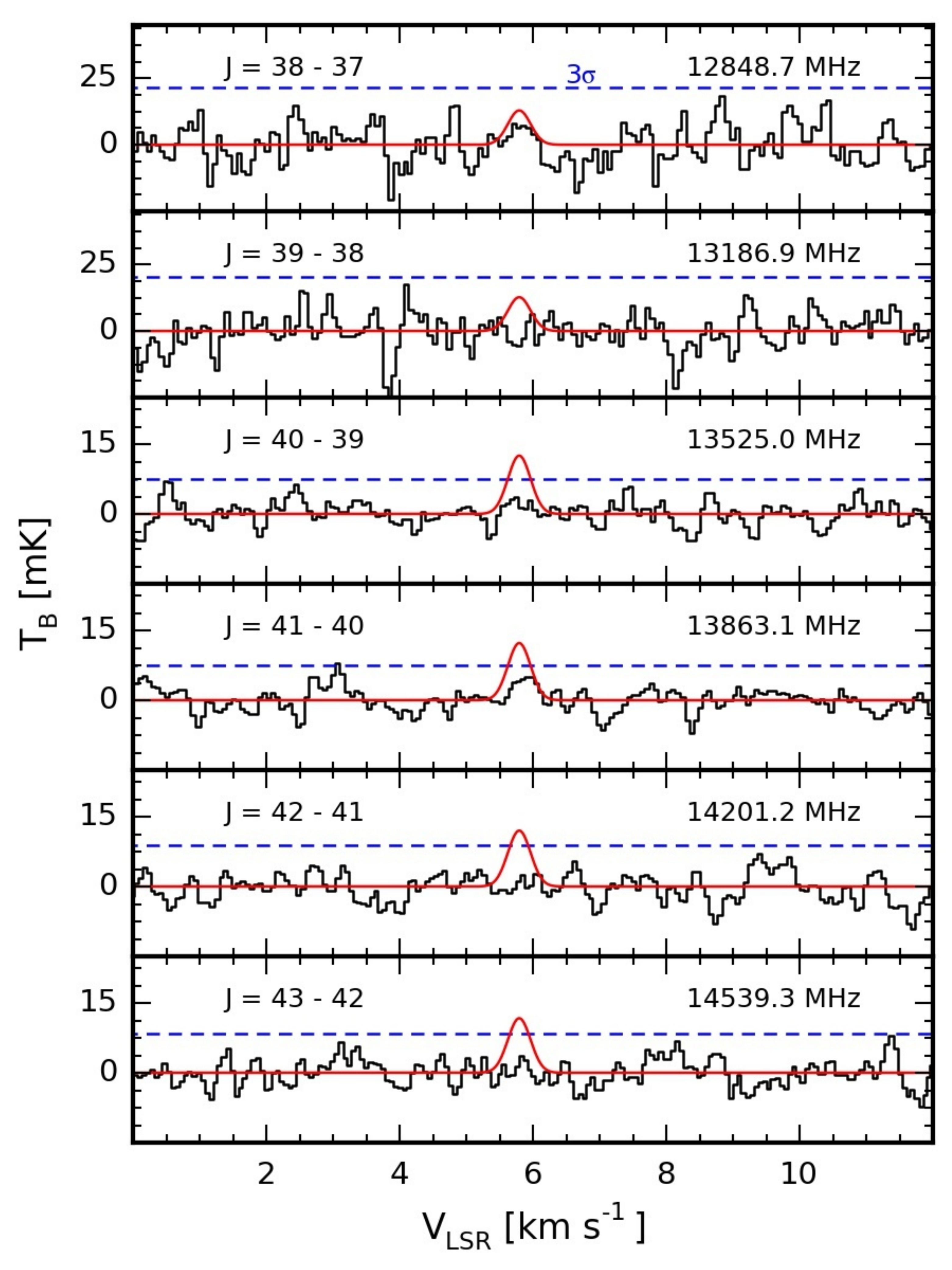}
\caption{{\small Observations of HC$_{11}$N rest frequencies with overplot of predictions from the \protect\cite{Bell_1997} column density in red. The 3$\sigma$ noise level for each spectrum is shown in blue.}}
\end{figure}

\begin{figure}
\centering
\includegraphics[width=\columnwidth]{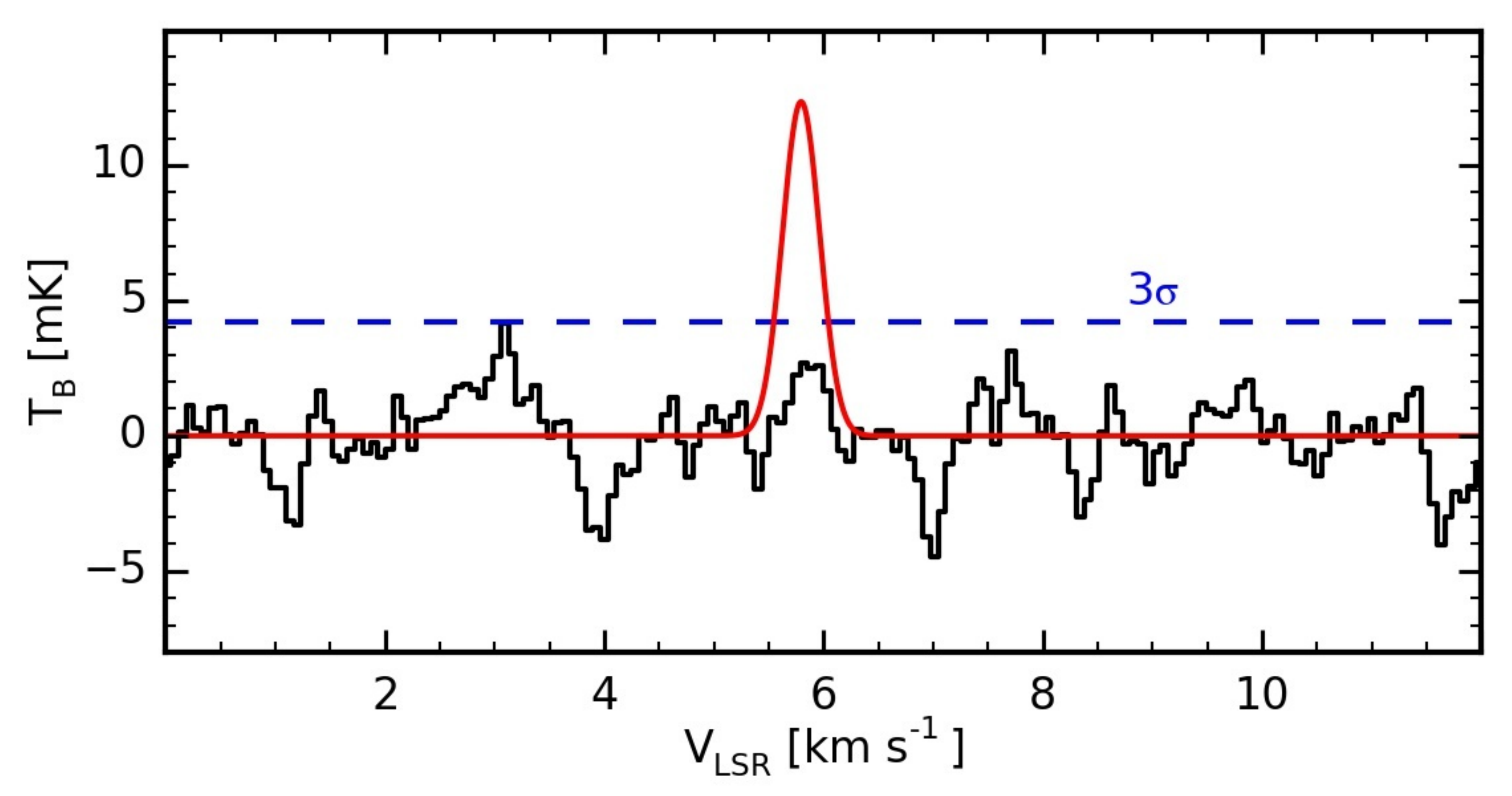}
\caption{{\small Weighted stacked spectra, with overplot of predictions from the \protect\cite{Bell_1997} column density in red. The 3$\sigma$ noise level is shown in blue.}}
\end{figure}

To quantitatively characterize an upper limit on the HC$_{11}$N column density, we used a Markov chain Monte Carlo (MCMC) code \citep{Foreman-Mackey_2013} to generate posterior probability distributions of the brightness temperature in each spectrum. The probability density function describes the range of possible brightness temperatures that are consistent with our observed data. As seen in Figure 2, the noise in the spectra are heavily correlated, due to the weighted convolution kernel used in re-gridding the spectra to correct for frequency smearing from Doppler tracking \citep[see Section 2.3 of][]{Langston_2007}. Thus, our calculated rms values for each spectrum are insufficient to describe the uncertainties in the data. To properly account for this correlation in the noise, we use a Gaussian Process Regression code\footnote{available at \url{https://github.com/dfm/george}} \citep{george} to construct a noise covariance matrix with a Mat\'ern 3/2 kernel. This approach more accurately describes our uncertainties, resulting in a larger, but more realistic upper limit. From these fits, we report the 95\% confidence level brightness temperature upper limits of each transition and their respective column density limits in Table 1.  Column density upper limits were calculated using the formalism described in \cite{Hollis_2004}, assuming optically thin emission, a rotational temperature of 10 K, and our measured HC$_9$N line-width of 0.37 km s$^{-1}$. 

Combining all of the spectral fits, we generate the total posterior probability distribution for column density, shown in Figure 4. As there is no clear detection of signal, the posterior is only useful in generating an upper limit, and we can exclude at a 95\% confidence level a column density greater than $\sim$9.4 $\times$ 10$^{10}$ cm$^{-2}$. Also shown in Figure 4 are the column density calculated from our chemical model (see Section 5), and the column density reported by \cite{Bell_1997}. To determine the compatibility of our results with those of \cite{Bell_1997}, we have additionally plotted a probability distribution for the \cite{Bell_1997} column density, calculated using the errors reported for their two detected line intensities. As seen in the figure, there is essentially no overlap between the two probability distributions. We therefore conclude that our observations are in disagreement with those of \cite{Bell_1997}.

\begin{figure}
\centering
\includegraphics[width=\columnwidth]{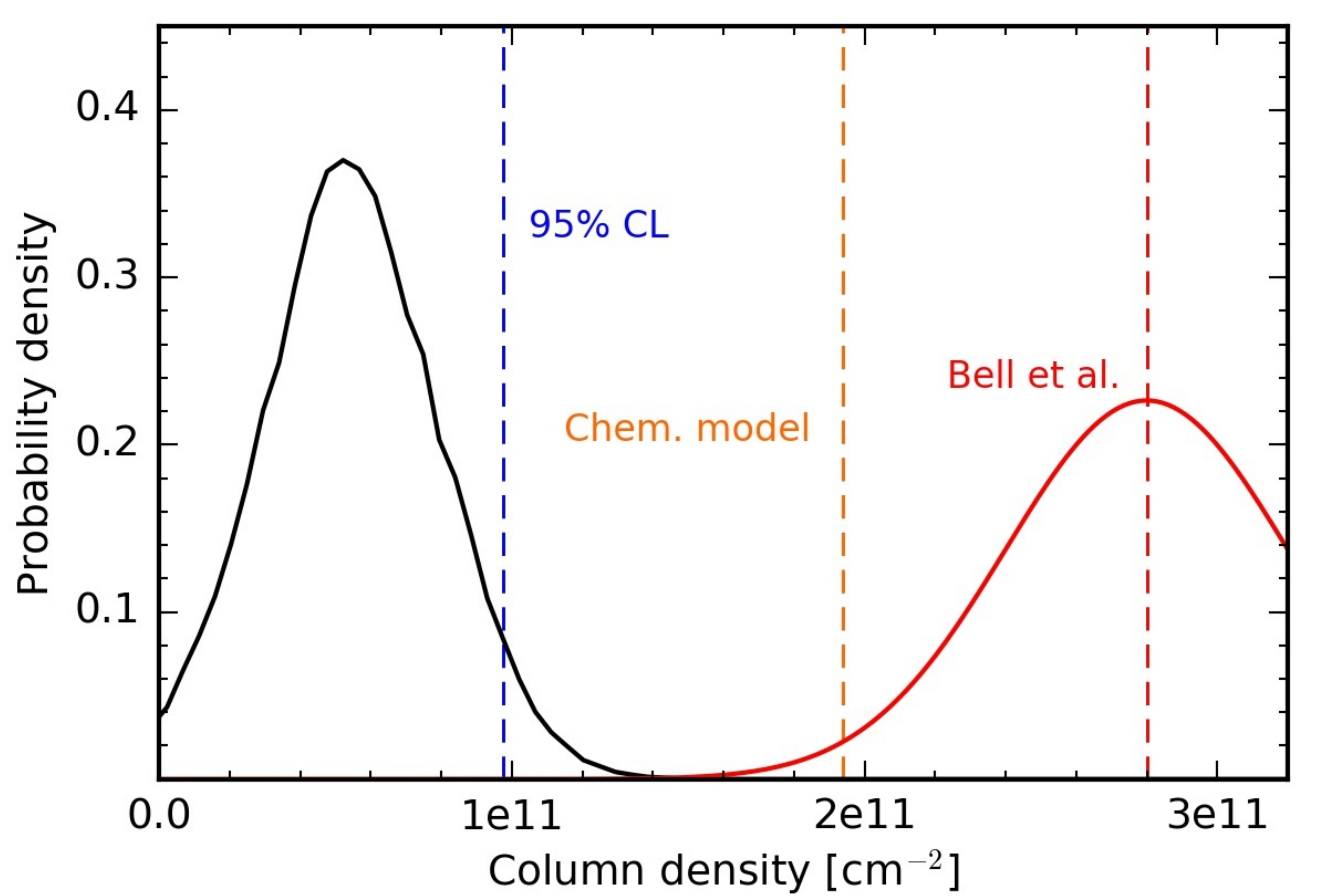}
\caption{{\small Posterior probability distribution of HC$_{11}$N column density derived from our MCMC fits to the individual spectra. The column density derived in \protect\cite{Bell_1997} and the predicted column density from our chemical model are also shown for comparison.}}
\end{figure}

In Figure 5, we plot our measured HC$_{9}$N and upper limit HC$_{11}$N column densities with literature cyanopolyyne column densities toward TMC-1 \citep{Bell_1997, Kaifu_2004, Kalenskii_2004, Remijan_2006, Langston_2007, Cordiner_2013}. Only one recent value is shown for the optically thick species HC$_3$N, as many literature HC$_3$N column densities are calculated assuming optically thin emission and are artificially low.  From these literature column densities, we have calculated a best fit linear trend, plotted as a black dashed line.  Our observed HC$_9$N column density is in good agreement with this trend, but our upper limit for HC$_{11}$N significantly deviates both from the \cite{Bell_1997} observations and the trend, suggesting additional cyanopolyyne chemistry at sizes larger than HC$_{9}$N.

\begin{figure*}
\centering
\includegraphics[width=0.8\textwidth]{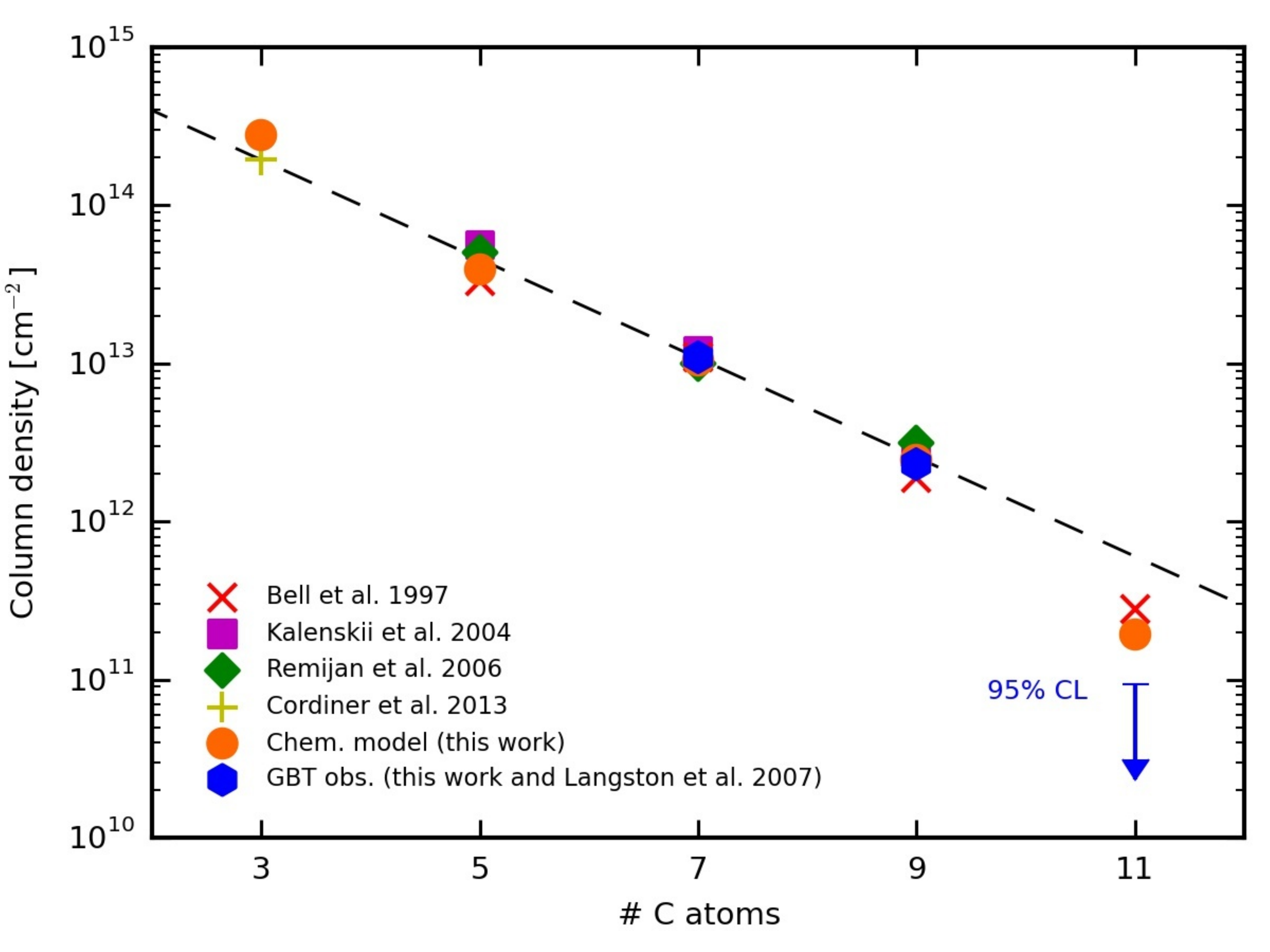}
\caption{{\small Cyanopolyyne column densities from the literature, along with our observations and chemical modeling results.}}
\end{figure*}

\section{Possibility of systematic errors}
We have shown that HC$_{11}$N emission is absent from the present GBT observations, in conflict with the column density reported by \cite{Bell_1997}. It is possible, however, that our analysis of the GBT observations is flawed due to systematic errors in either the reduced data or in our assumptions. Here we investigate four possible sources of error in our analysis: contamination in the ``off" position during the nodding observations, an inaccurate assumption of the beam filling factor, antenna temperature calibration uncertainties, {and the validity of an LTE analysis.

\subsection{``Off" position contamination}
The GBT observations were taken in a nodding mode, and in the data reduction the ``off" position data were subtracted from the ``on" position data. Any molecular emission from the ``off" position would weaken the measured line strengths, possibly dramatically decreasing their intensity.  This is especially a concern in an extended source such as TMC-1, as the angular offset of the two beams of the Ku-band receiver is relatively small (330\arcsec).  While TMC-1 extends over 10\arcmin~along its principal axis, cyanopolyyne emission arises from a more compact region, spanning roughly 6.0\arcmin~x 1.3\arcmin~\citep{Churchwell_1978, Olano_1988, Bell_1997}. Our 5.5\arcmin~nodding offset is then likely sufficient to prevent contamination, placing the ``off" beam a safe distance away from the emission region.  By examining the raw ``on" and ``off" spectra for HC$_9$N (taken simultaneously with the HC$_{11}$N spectra) and assuming that HC$_9$N would trace the same emission region as HC$_{11}$N, we are able to rule out this possible source of systematic error (Figure 6).

\begin{figure}
\centering
\includegraphics[width=\columnwidth]{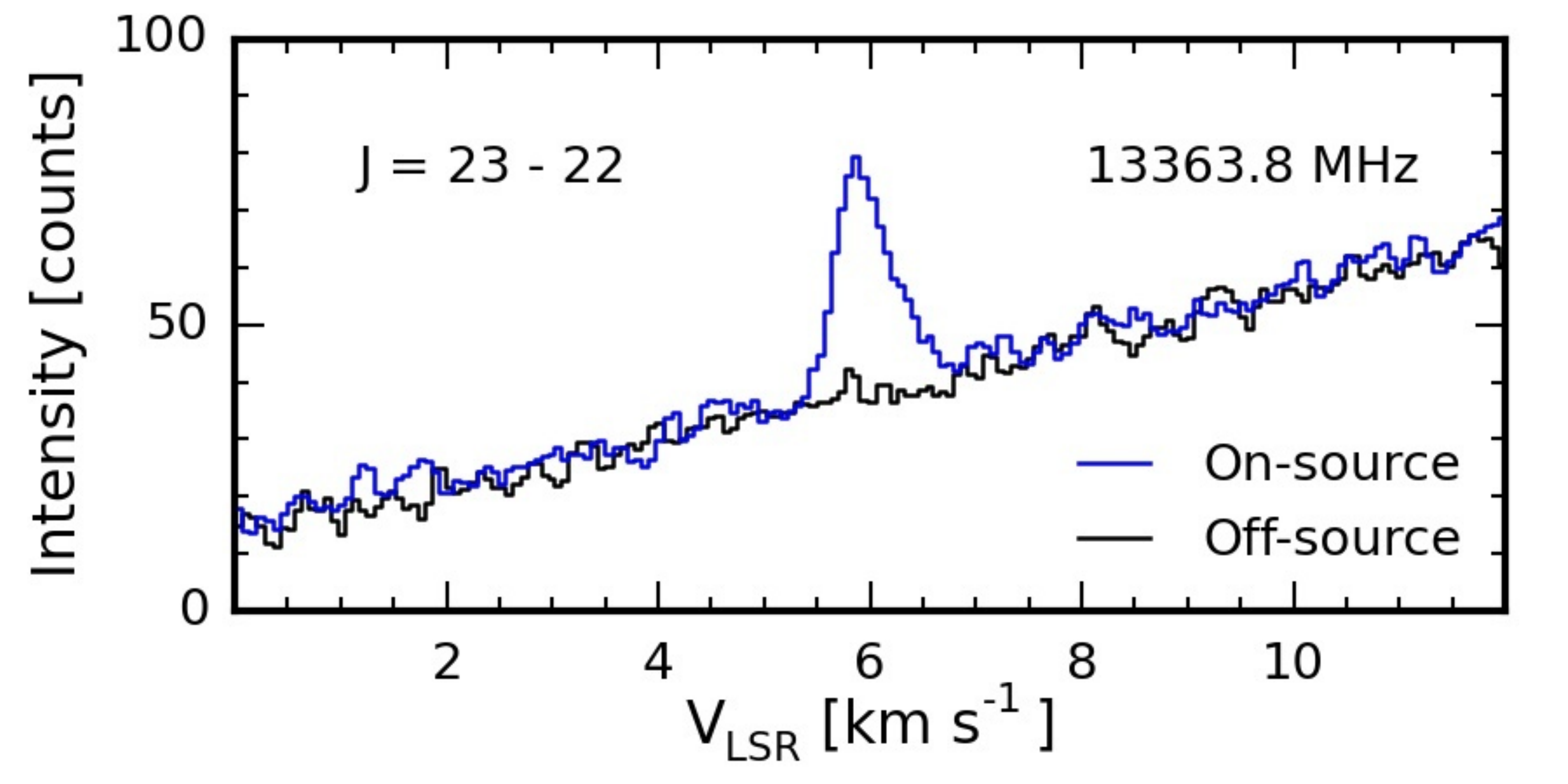}
\caption{{\small Raw on-source and off-source nodding spectra of the HC$_9$N J=23--22 transition, shown in blue and black respectively. As there is no ``off" source to calibrate the ``off" data against, an intensity unit of counts is used, and a scalar value is subtracted from the ``on" data to align them. No HC$_9$N emission is observed in the off-source spectrum.}}
\end{figure}

\subsection{Beam filling factor}
The NRAO 140 ft telescope and the GBT have beams that differ in size by approximately a factor of 2 at a given frequency.  Comparison of derived column densities therefore rests upon an accurate assessment of the beam filling factor of the source emission region. In their column density calculation, \cite{Bell_1997} assume that all cyanopolyyne emission arises from the previously mentioned 6.0\arcmin~x 1.3\arcmin~region, yielding a beam-filling factor of 0.54 for their 2.4\arcmin~beam. The mapping results of \cite{Hirahara_1992}, \cite{Pratap_1997}, and \cite{Fosse_2001} show that HC$_3$N emission traces a region approximately this size and shape, confirming that the \cite{Bell_1997} beam-filling factor estimate is reasonable. We can then safely assume that the 1.2\arcmin~GBT beam is completely filled.

\subsection{Antenna temperature calibration}
The antenna temperatures for our observations were calibrated using noise diodes (see \cite{Langston_2007} for details), while those from \cite{Bell_1997} were recorded on the T$_{A}^{*}$ scale, presumably calibrated using a chopper wheel \citep[e.g.][]{Ulich_Haas_1976}. Antenna temperature calibration uncertainty is often reported at levels of 10 -- 30$\%$, and might possibly introduce a relative bias between the two measurements. Although we cannot directly constrain the relative uncertainty due to either the antenna temperature calibration or beam filling factor assumptions, we can quantify the uncertainty in derived column densities, which is the compared quantity between the two observations.

For optically thin lines, a bias in antenna temperature will correspond linearly to a bias in derived column density, as would a change in the beam filling factor assumption. Since the smaller cyanopolyyne lines have numerous observations \citep[e.g.][]{Bell_1997, Kalenskii_2004, Remijan_2006, Langston_2007}, we can use the scatter in the derived column densities to estimate how biased a given set of observations might be. Without including the results of \cite{Bell_1997}, the scatter for the HC$_{5}$N, HC$_{7}$N, and HC$_{9}$N column densities shown in Figure 5 is 9$\%$. Including the \cite{Bell_1997} column densities brings this scatter to 16$\%$, as they are slightly systematically low compared to literature averages. The column densities reported here and in \cite{Langston_2007} are well within this scatter, suggesting that there is unlikely to be a greater than $\sim$20$\%$ column density bias between our observations and \cite{Bell_1997}. As \cite{Bell_1997} column densities are systematically low, this might signify a small error in either their antenna temperature calibration or assumed beam filling factor. If this was to be corrected, however, it would increase their HC$_{11}$N column density and create more conflict between the two datasets. Even so, to be conservative we compare our derived column density with that of \cite{Bell_1997} shifted down by 20$\%$, and we find a \textless1$\%$ chance that the two datasets are compatible. We additionally note that our derived HC$_{9}$N/HC$_{11}$N ratio is quite different from that of \cite{Bell_1997} (\textgreater24 vs 6.8), which should be independent of any absolute calibration uncertainty. Thus we conclude that uncertainties in calibration or beam filling factor are unlikely to be able to explain discrepancies between the two sets of observations.

\subsection{Validity of LTE}
In our line strength predictions and calculations of column densities we have assumed LTE excitation of the cyanopolyynes in TMC-1, as was assumed in \cite{Bell_1997, Bell_1998, Ohishi_1998, Kaifu_2004, Langston_2007}. The critical densities of HC$_9$N and HC$_{11}$N are not well known, as their collisional cross-sections have not been measured, but can be estimated by scaling the cross section of HC$_3$N \citep{Green_1978, Avery_1979, Snell_1981}. The change to the cross-section is estimated as a linear scaling proportional to the size of the molecule, yielding an HC$_{11}$N cross section $\sim$3$\times$ larger than that of HC$_3$N. In reality, the cross section is likely significantly larger, but the linear scaling provides a good upper limit estimate of the critical density. From this scaled cross-section ($\sim$3$\times$10$^{-14}$cm$^{2}$), an assumption of a kinetic gas temperature of 10K, and A$_{ij}$ values from the Splatalogue database, we estimate a critical density between 300 and 800 cm$^{-3}$ for the HC$_{11}$N transitions covered by our observations. Similarly, we estimate a critical density between 400 and 900 cm$^{-3}$ for our observed HC$_9$N transitions. These critical densities are significantly smaller than the gas density of TMC-1, $2\times10^4$ cm$^{-3}$ \citep{Liszt_Ziurys_2012, Hincelin_et_al_2011}, suggesting that the observed transitions are fully thermalized and our assumption of LTE is reasonable.  Future work on the collisional cross sections of larger cyanopolyynes may allow for more detailed statistical equilibrium calculations and a more careful analysis of the observed linear abundance trend in cyanopolyyne column densities.

\subsection{Possible correlator artifacts}
Finally, we turn to the possibility that the \cite{Bell_1997} observations are spurious. The HC$_{11}$N features observed by \cite{Bell_1997} could be attributed to frequency switching artifacts from the correlator, which they mention as a possible explanation for several nearby U-lines. The small-offset frequency switching method used by \cite{Bell_1997} in particular may be susceptible to this problem; telluric lines and local interference can be removed using a short off-source integration, but this would not remove the effects of weak correlator artifacts \citep{Bell_1991, Bell_1993}. \cite{Bell_1997} mention that they utilize two sets of observations with HC$_{11}$N in different filter banks to reduce these effects, but it is unclear how effective this is, especially in conjunction with their LINECLEAN method of removing frequency-switched images \citep{Bell_1997b}. We do not observe U-lines of similar intensity in our position switched data (Figures 1 and 2), and no known molecular lines correspond to the U-lines recorded in \cite{Bell_1997}. Additionally, the noise in the \cite{Bell_1997} spectra is correlated on roughly similar scales to the characteristic line-width of TMC-1 (0.4 km s$^{-1}$), which could explain their observed line-widths if the HC$_{11}$N features and U-lines were artifacts.

In short, we are confident in the quality of the observations presented here and in \cite{Langston_2007}, as our measured column densities agree quite well with previous observations. We are therefore also confident in our non-detection of HC$_{11}$N in this dataset, and the derived HC$_{9}$N/HC$_{11}$N column density ratio that contradicts the previously suggested log-linear abundance trend \citep{Remijan_2006}.  Although our analysis suggests that our observations are in conflict with those of \cite{Bell_1997}, we can not conclusively show whether this is due to measurement and calibration errors, or whether the \cite{Bell_1997} detection is spurious. In either case, the deviation of the HC$_{11}$N abundance from the log-linear trend requires further chemical investigation.

\section{Chemical modeling}
\subsection{Modeling code and updated reactions}

As shown in Figure 5, our observations in combination with literature values suggest that although the smaller cyanopolyynes have a log-linear abundance trend, HC$_{11}$N is clearly an outlier. To investigate possible chemical explanations of this, we have modeled the chemistry of HC$_{11}$N and other smaller cyanopolyynes in TMC-1. We used the deterministic gas-grain rate equation model Nautilus \citep{nautilus_code} with the 2014 KIDA network of chemical reactions \citep{kida_network1,kida_network2}, updated to include reactions for HC$_{11}$N and several related species (see Table 2 and the online supplementary data).  The physical conditions used in the model were chosen to be representative of TMC-1, i.e. a kinetic temperature of 10 K and a gas density of $2\times10^4$ cm$^{-3}$, as were the initial elemental abundances \citep{Liszt_Ziurys_2012, Hincelin_et_al_2011}.

\begin{table*}
\begin{center}
\small
\begin{threeparttable}[b]
\caption{Reactions Added to Network$^{\mathrm{a}}$}
\begin{tabular}{+l^c^c^c}
\toprule
	\multicolumn{1}{c}{Reaction} & $\alpha^{b}$ & $\beta^{b}$  & $\gamma^{b}$  \\ 
	\toprule
	HC$_{11}$N + photon $\rightarrow$ C$_{11}$HN$^+$ + e$^-$        & $2.0\times 10^{-10}$  & $0.0$	    & 2.5	\\
	CN + C$_{10}$H$_2$ $\rightarrow$ H + HC$_{11}$N                 & $2.7\times 10^{-10}$  & $-0.5$    & 19	\\
	HC$_{11}$N + H$^+$ $\rightarrow$ H + C$_{11}$HN$^+$             & $4.9\times 10^{-08}$  & $-0.5$    & 0	    \\
	HC$_{11}$N + H$_3^+$ $\rightarrow$ H$_2$ + C$_{11}$H$_2$N$^+$   & $2.7\times 10^{-08}$  & $-0.5$    & 0	    \\
	HC$_{11}$N + HCO$^+$ $\rightarrow$ CO + C$_{11}$H$_2$N$^+$      & $1.7\times 10^{-08}$  & $-0.5$    & 0     \\
	\bottomrule
\end{tabular}
\begin{tablenotes}
\item[a] Table 2 is published in its entirety in the electronic edition of \textit{MNRAS}, A portion is shown here for guidance regarding its form and contents. 
\item[b] Parameter units depend on the type of reaction and rate-coefficient formula
\end{tablenotes}
\end{threeparttable}
\end{center}
\end{table*}

In the model, our added reactions for HC$_{11}$N follow those of the smaller cyanopolyynes included in the network. The formation chemistry of cyanopolyynes and other unsaturated carbon-chain species is dominated by gas-phase reactions. Grain-surface contributions are negligible, as unsaturated carbon chains rapidly react with hydrogen on grain surfaces, producing saturated hydrocarbons. Unsaturated carbon-chain molecules are often formed via reactions between smaller carbon-chains through the addition of one or more carbons to the backbone. One such reaction, which is a major formation route for cyanopolyynes \citep{Fukuzawa_1998}, is 

\begin{equation}
\mathrm{CN} + \mathrm{C_{\textit{n}-1}H_2} \rightarrow \mathrm{H} + \mathrm{HC_\textit{n}N}.
\end{equation}

\noindent
Cyanopolyynes can also form via reactions involving precursors with the same number of carbons, two important examples of which are:

\begin{equation}
	\mathrm{H_2C_\textit{n}N^+} + \mathrm{e^-} \rightarrow \mathrm{H} + \mathrm{HC_\textit{n}N}
\end{equation}

\begin{equation}
        \mathrm{H_3C_\textit{n}N^+} + \mathrm{e^-} \rightarrow \mathrm{H_2} + \mathrm{HC_\textit{n}N}.
\end{equation}

\noindent
Finally, the destruction of cyanopolyynes is dominated by reactions with ionic species. These have the general form

\begin{equation}
	\mathrm{HC_\textit{n}N} + \mathrm{X^+} \rightarrow \mathrm{products}
\end{equation}

\noindent
where X$^+$ is some ionic species, e.g. C$^+$, H$^+$, H$_3^+$, or HCO$^+$. Ionized carbon is a particularly important co-reactant in many of the destruction pathways, and is formed in the model mainly via the dissociation of CO by cosmic-ray ionized helium \citep{Rimmer_2012}, i.e.

\begin{equation}
  \mathrm{CO} + \mathrm{He}^+ \rightarrow \mathrm{C^+} + \mathrm{O} + \mathrm{He}.
\end{equation}

The rate coefficients for most reactions in the code are calculated using the Arrhenius-Kooij formula,

\begin{equation}
	k(T) = \alpha \left(\frac{T}{300\;\mathrm{K}}\right)^{\beta} \;\mathrm{exp}^{-\gamma / T}
\end{equation}

\noindent
where $\alpha$ is a temperature independent pre-exponential factor, $\beta$ governs the temperature dependence, and $\gamma$ is the energy barrier. Many of the reactions are difficult, if not impossible, to study in the laboratory. Thus in the absence of experimental values, rate coefficients were estimated via extrapolation from similar reactions involving smaller carbon-chain species or via a standard formula such as the Langevin rate for ion-dipole reactions. Our added reactions and their corresponding values of $\alpha$, $\beta$, and $\gamma$ are listed in Table 2.

\subsection{Ion-dipole effects}

Reactions between ions and neutral molecules with a dipole moment are often both barrier-less and inversely temperature dependent, i.e. $\gamma=0$ and $\beta<0$ \citep{Clary_2008}, meaning that they are particularly efficient in cold regions such as TMC-1.  The rate coefficients for these reactions are proportional to the dipole moment of the neutral reactant molecule, and the 2014 KIDA network has the advantage of accounting for this phenomenon.

For the reactions mentioned above, the rate coefficients will increase with cyanopolyyne length, due to the enhancing effect of both a larger physical cross-section and dipole moment. For example, the ratio of rate coefficients for the destruction reactions of HC$_{11}$N and HC$_3$N with the ion HCO$^+$, $k_\mathrm{HC_{11}N}/k_\mathrm{HC_3N}$, is $\sim$4.75. Thus larger species, though less abundant, are more reactive.  This should add a small non-linear contribution to the cyanopolyyne abundance trend, which will become more pronounced in larger molecules such as HC$_{11}$N.

\subsection{Comparison of model results and observations}

The chemical model was run to a final time of 10 Myr, and we derive a best fit time of $\sim$1.6 Myr by comparing cyanopolyyne abundance ratios from literature averages and the model (Figure 7). This age is similar to previous estimates of the chemical age of TMC-1 \citep{Brown_1990, Ruffle_2001}. The chemical model abundances from 1.6 Myr are given in Table 3. To scale these abundances to column densities, we fit a proton column density of 1.3 $\times$10$^{23}$ cm$^{-2}$. Assuming that molecular hydrogen is the dominant proton carrier, this roughly corresponds to a molecular hydrogen column density of 7 $\times$10$^{22}$ cm$^{-2}$, within an order of magnitude of other estimates \citep{Suutarinen_2011}. For HC$_{11}$N, the predicted abundance relative to the proton density is $1.4\times10^{-12}$, yielding a column density of 1.9 $\times$10$^{11}$ cm$^{-2}$.  While this column density falls below the predicted linear trend \citep{Remijan_2006} and the observed value of $2.8\times10^{11}$ cm$^{-2}$ \citep{Bell_1997}, it is still above our 95\% confidence level upper limit column density of $9.4\times10^{10}$ cm$^{-2}$ (Figures 4 \& 5). Thus, although the additional consideration of ion-dipole effects in our modeling relieves some tension with the data by introducing a non-linear contribution to the abundance trend, there may be additional chemistry unaccounted for by the model.

\begin{table*}
\begin{center}
\small
\begin{threeparttable}[b]
\caption{Cyanopolyyne abundances and calculated column densities}
\begin{tabular}{+c^c^c^c}
\toprule
	            & Abundance$^{a}$       & Column Density$^{b}$          & Column Density$^{c}$          \\
	Molecule    & Chemical Model        & Chemical Model                & Observations                  \\
	            & ($\times$10$^{-12}$)  & ($\times$10$^{11}$ cm$^{-2}$) & ($\times$10$^{11}$ cm$^{-2}$) \\ 
	\otoprule
	HC$_{3}$N   &	2060  &	2770	&	1950	\\
	HC$_{5}$N   &	291   &	392 	&	452	    \\
	HC$_{7}$N   &	78    &	105	    &	110	    \\
	HC$_{9}$N   &	18    &	25	    &	25	    \\
	HC$_{11}$N  &	1.4   &	1.9	    &	2.8$^d$ \\
	\bottomrule
\end{tabular}
\begin{tablenotes}
\item[a] Relative to the proton density, $n_\mathrm{p} = n_\mathrm{H} + 2n_\mathrm{H_2}$.
\item[b] Calculated from chemical model abundances assuming n$_{H_{2}}$ = 7 $\times$ 10$^{22}$ cm$^{-2}$.
\item[c] Calculated from average of all literature measurements presented in Figure 5 (see Section 3).
\item[d] Observed column density from \cite{Bell_1997}. Our 95\% confidence level upper limit column density is 9.4$\times$10$^{10}$ cm$^{-2}$
\end{tablenotes}
\end{threeparttable}
\end{center}
\end{table*}

\begin{figure}
\centering
\includegraphics[width=\columnwidth]{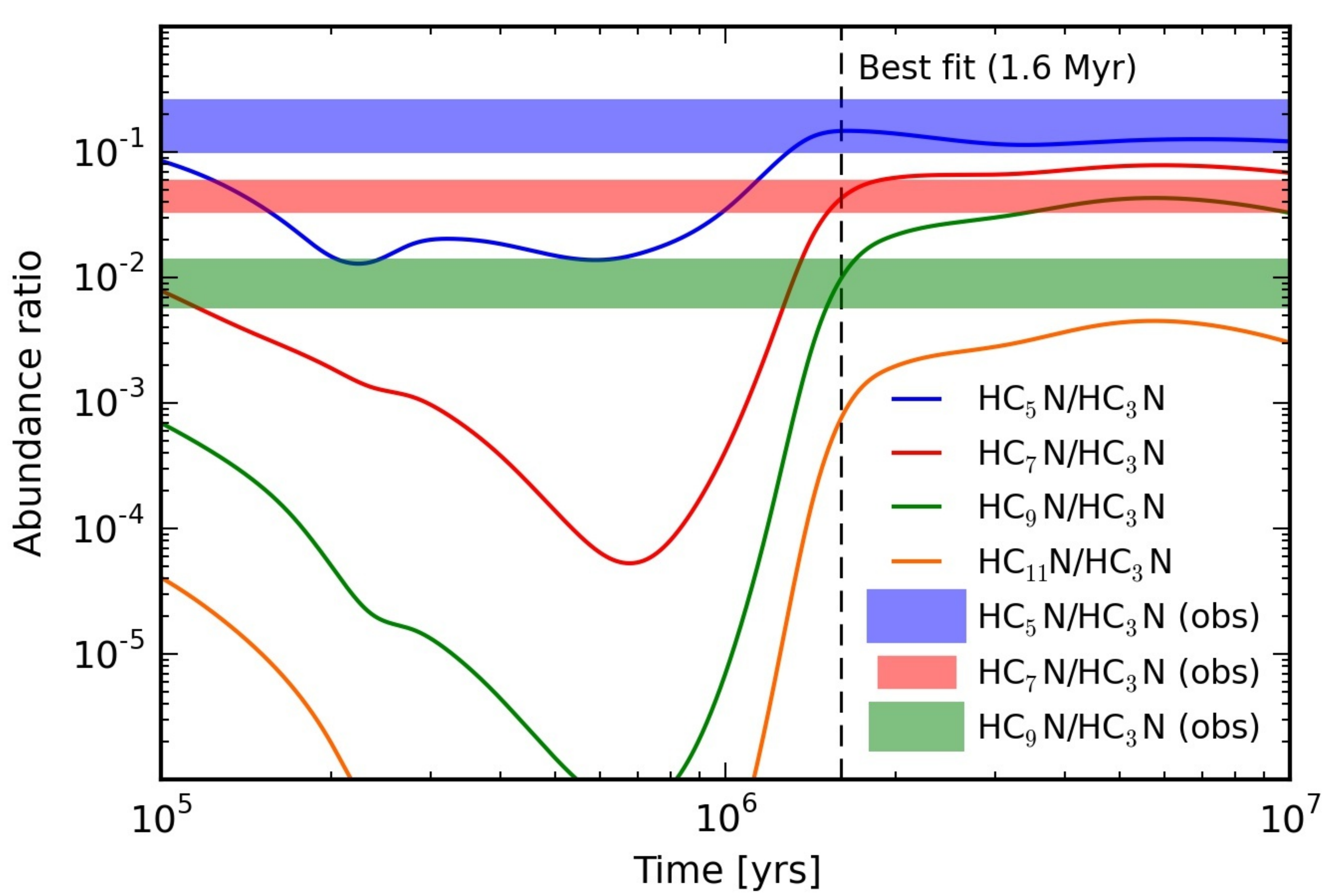}
\caption{{\small Ratios of cyanopolyyne abundances in our chemical model, with observational column density ratios overplotted. The spread in each observational column density ratio is shown by the width of the line.}}
\end{figure}

\section{Discussion}

As suggested by \cite{Bell_1997} and \cite{Remijan_2006}, it has been previously assumed that the chemistry of long (\textgreater 10 atoms) carbon-chain species would follow the general pattern of the smaller ones, and that it should be possible to detect not only HC$_{11}$N, but HC$_{13}$N as well. This is the basic assumption behind the added HC$_{11}$N reactions in our chemical model. Our observations, however, suggest additional chemistry for cyanopolyynes, and possibly other carbon-chain species, larger than HC$_9$N.

Our low upper limit on the abundance of HC$_{11}$N suggests that there may be an unknown destruction mechanism for either HC$_{11}$N or one of its precursors, such as C$_{11}$H$_2$N$^+$. Although destruction through reaction with ions is considered in our chemical model, isomerization reactions such as cyclization were not included. Cyclization of large cyanopolyynes may be supported by studies of pure carbon clusters, i.e. C$_n$, which are analogous to cyanopolyynes in their carbon-chain structure (see review article by \cite{Weltner_van_Zee_1990}).  Experiments show that carbon clusters up to 9 atoms exist mainly in a linear form, while 10 atom clusters are a mixture of linear and monocyclic isomers, and clusters of 11 or more atoms exist purely as cyclic structures \citep{von_Helden_1993}. This behavior is easily understood, as cyclic isomers become more thermodynamically stable than their linear counterparts at larger sizes \citep{van_Orden_1998}. Additionally, at larger sizes the number of possible isomers rapidly increases and the barriers to isomerization reactions shrink, making spontaneous isomerization more likely. These properties should generally extrapolate from carbon clusters to cyanopolyynes, and thus at sizes larger than 11 atoms (i.e. larger than HC$_{9}$N), the cyclic forms of long carbon-chains could become a ``sink", depleting abundances of linear forms \citep{Travers_1996, McCarthy_2000}.

The products of this isomerization would depend on the parent species, and might contain both cyclic and linear components, resembling known ring-chain molecules such as C$_7$H$_2$, C$_9$H$_2$, HC$_4$N, and HC$_6$N \citep{McCarthy_1997, McCarthy_1998, McCarthy_1999}. None of these molecules have yet been found in their ring-chain forms toward interstellar sources \citep{Cernicharo_2004}, but this may be related to their small size, as the cyclization reactions would be more prevalent for the larger polyynes.  Our observations suggest that more laboratory work should be conducted to investigate the rotational spectra of larger ring-chain species, and that systematic searches in survey data toward TMC-1 or IRC-10216 may prove fruitful. Other sources which might host these molecules are the ``carbon-chain-producing regions" such as Lupus-1A \citep{Hirota_2006, Hirota_2009, Sakai_2010}.

\section{Summary}

The results of our analysis of GBT observations of cyanopolyynes toward TMC-1 can be summarized as follows:

\begin{enumerate}
\item Six transitions of HC$_9$N were detected; we derive a column density of 2.3$\pm$0.2 $\times$ 10$^{12}$ cm$^{-2}$ and a rotational temperature of 10$\pm$2 K, consistent with previous measurements.
\item Six transitions of HC$_{11}$N were covered by the observations, but none were detected above 2$\sigma$ significance.
\item We place a 95\% confidence level upper limit on the HC$_{11}$N column density of 9.4 $\times$ 10$^{10}$ cm$^{-2}$, in conflict with the \cite{Bell_1997} reported column density of 2.8 $\times$ 10$^{11}$ cm$^{-2}$. This is also a significant deviation from previous predictions of a log-linear trend in cyanopolyyne abundances.
\item Our chemical model produces a depleted HC$_{11}$N abundance when the effects of ion-dipole enhancement are included, but is still not able to reproduce the observed upper limit column density.
\item Supported by studies of carbon clusters, cyclization reactions may play a role in the depletion of HC$_{11}$N. The products of these reactions should be pursued as candidate interstellar molecules in future laboratory and astronomical studies.
\end{enumerate}

\section*{Acknowledgements}

We would like to thank M. McCarthy, A. Vanderburg, and B. Montet for productive discussion and helpful comments on the manuscript. We thank two anonymous referees for providing comments that greatly improved the quality of the manuscript. RAL gratefully acknowledges support from a National Science Foundation Graduate Research Fellowship. CNS wishes to thank the National Science Foundation for supporting the Astrochemistry program at the University of Virginia. BAM and PBC are grateful to G. A. Blake for his support.  The National Radio Astronomy Observatory is a facility of the National Science Foundation operated under cooperative agreement by Associated Universities, Inc.




\bibliographystyle{mnras}

\bsp	
\label{lastpage}
\end{document}